# *Rotating Spacetime Modulation: Topological Phases and Spacetime Haldane Model*

*João C. Serra, Mário G. Silveirinha***

*University of Lisbon–Instituto Superior Técnico and Instituto de Telecomunicações,
Avenida Rovisco Pais, 1, 1049-001 Lisboa, Portugal*

**Abstract**

Topological photonics has recently emerged as a very general framework for the design of unidirectional edge waveguides immune to back-scattering and deformations, as well as other platforms that feature extreme nonreciprocal wave phenomena. While the topological classification of time invariant crystals has been widely discussed in the literature, the study of spacetime crystals formed by time-variant materials remains largely unexplored. Here, we extend the methods of topological band theory to photonic crystals formed by "inclusions" that are subject to a spacetime rotating-wave modulation that imitates a physical rotating motion. By resorting to an approximate non-homogeneous effective description of the electromagnetic response of the inclusions, it is shown that they possess a bianisotropic response that breaks the time-reversal symmetry and may give rise to non-trivial topologies. In particular, we propose an implementation of the Haldane model in a spacetime modulated photonic crystal.

---

**\*** To whom correspondence should be addressed: E-mail: *mario.silveirinha@co.it.pt*

# I. Introduction

Topological photonics provides a very general framework to classify different classes of photonic crystals with band gaps, and to design unidirectional edge-type channels immune to back-scattering formed by photonic crystals with different topologies [1-4]. Traditional photonic crystals are material structures with a periodic space modulation of the refractive index [5, 6]. In recent years, time has been explored as a new degree of freedom in material design, expanding the photonic crystal notion also to structures presenting time or spacetime modulations [7-14]. In particular, the spacetime modulation has emerged as a promising new paradigm to break the electromagnetic reciprocity and achieve giant electrically-tunable nonreciprocal responses without using a magnetic bias [15-18]. It is also worth pointing out that even a static electric bias can originate in some conditions a nonreciprocal response [19-21].

It is known that reciprocal systems are necessarily characterized by trivial Chern topologies [22-25]. However, it is not sufficient to break the reciprocity (time-reversal symmetry $T$) to achieve non-trivial topologies, as there are other symmetries such as mirror ($P_x$) and parity-time ($PT$) symmetries that also lead to trivial Chern indices. Typically, nontrivial topological phases are engineered with the help of a static magnetic bias, exploiting either an electric gyrotropic or a magnetic gyrotropic material response [22-24, 26-28]. The use of an external magnetic bias leads to bulky components, which may not be very practical for some nanophotonic applications. For this reason, metamaterials with time or spacetime modulations may be interesting alternatives to create magnetic-free non-reciprocal systems.

Photonic crystals with spacetime travelling-wave modulations, e.g., $\varepsilon = \varepsilon(x - vt)$ with $v$ the modulation speed, have recently attracted considerable attention as they can be used to engineer a synthetic (electronic) motion that imitates the translational motion of a material structure [12, 13, 14, 29]. Different from moving systems, in a spacetime crystal the modulation speed $v$ is not bounded by the speed of light, and hence superluminal regimes are physically admissible [12, 13, 14, 29].

Heuristically, a nontrivial topological phase must be associated with some internal angular momentum of the material [4, 30], analogous to the angular momentum imparted by a static magnetic bias through the generation of cyclotron orbits. This property suggests that the most suitable spacetime modulation is related to a physical rotation. Motivated by this idea, here we consider two crystals that consist of periodic arrangements of "ring resonators" subject to a spacetime rotating-wave modulation of the type $\varepsilon = \varepsilon(\varphi - \Omega_0 t)$ and $\mu = \mu(\varphi - \Omega_0 t)$. This modulation is the angular equivalent of the linear travelling-wave modulation. By resorting to an approximate effective medium description of the resonators, it is shown that they effectively behave as nonuniform materials with a bianisotropic response that breaks simultaneously the $P_x$, $T$ and $PT$ symmetries and generates a non-trivial topology. Thereby, we introduce a new paradigm to generate nontrivial topological phases by exploiting spacetime modulations in a 2D crystal. It should be noted that different topological aspects of time-varying structures have been discussed in previous works in the context of time crystals [8], of systems with synthetic frequency dimensions [31-34] and of higher-order topological insulators [35-36].

The paper is organized as follows. In section II, we develop an effective medium model for a spacetime modulated ring, which is the basic building block of the spacetime crystals. Section III describes the theoretical methods used to compute the band structure and the topological phases of the relevant spacetime crystals. Section IV presents two cases of interest that illustrate how the spacetime modulation of different sublattices of ring resonators affects the Chern index: first, we consider a honeycomb lattice; second, we propose a spacetime analogue of the Haldane model. Finally, the main results are summarized in section V.

## II.    Effective model for a spacetime modulated ring resonator

Let us consider a cylindrical spacetime modulated ring resonator with internal and external radii $R_{in}$ and $R_{ext}$ centered at the origin. This resonator will be used in the following sections as the basic building block of different spacetime crystals. Related ring resonators were studied in Refs. [16, 17] in a different context. The material parameters of the resonator are of the form $\varepsilon = \varepsilon(\varphi - \Omega_0 t)$ and $\mu = \mu(\varphi - \Omega_0 t)$, with $\Omega_0$ the modulation angular frequency as illustrated in Fig. 1. It is supposed that $\varepsilon(\varphi') = \varepsilon(\varphi' + \Delta\varphi)$ and $\mu(\varphi') = \mu(\varphi' + \Delta\varphi)$ with $\Delta\varphi = 2\pi/N$, $N$ being the number of "unit cells" in the ring perimeter.

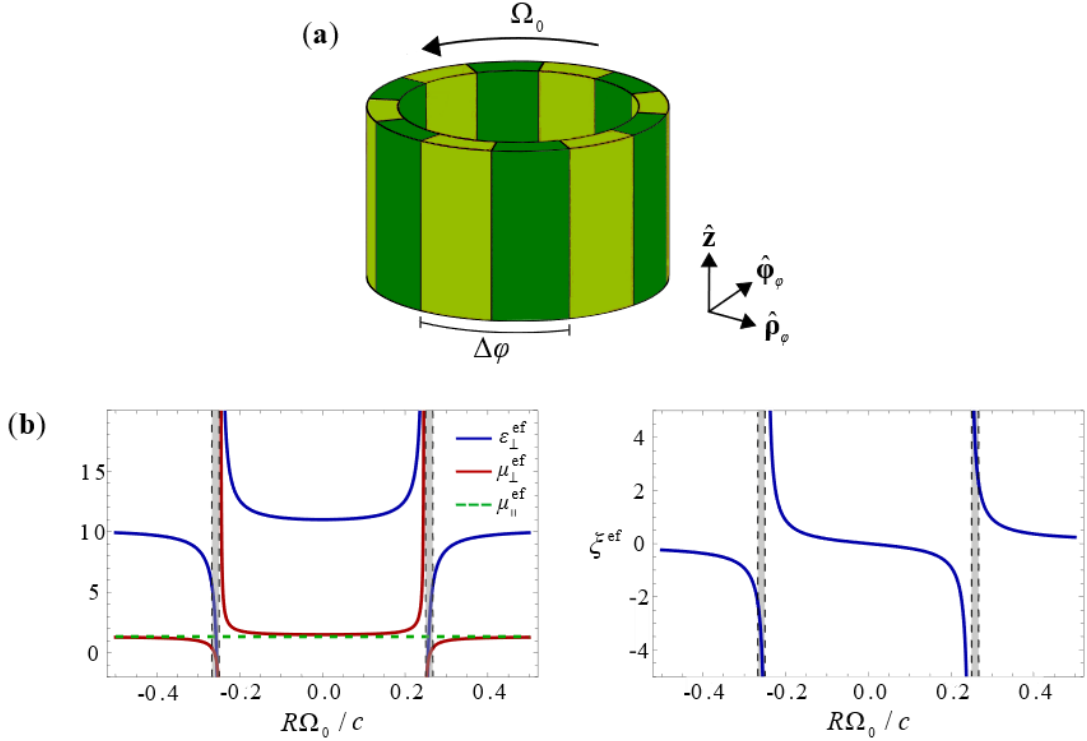

**Fig. 1 (a)** Geometry of a cylindrical ring resonator subject to a rotating spacetime modulation. **(b)** Relative effective parameters for TE modes as a function of the modulation angular frequency. The resonator is formed by layers with $\left(\varepsilon_{r,1}=14, \mu_{r,1}=1\right)$ and $\left(\varepsilon_{r,2}=8, \mu_{r,2}=2\right)$. The volume fraction of the two materials is identical. The sub-luminal/super-luminal thresholds are represented by vertical dashed black lines.

Following Ref. [29], a 1D crystal subject to a linear travelling-wave modulation behaves effectively in the long wavelength limit as an equivalent bianisotropic medium with a moving-type magneto-electric coupling. Provided the thickness of the ring resonator $\delta R = R_{ext} - R_{in}$ is much smaller than the average radius $R = \left(R_{ext} + R_{in}\right)/2$, the curvature effects may be neglected and the ring may be regarded as a 1D crystal. In other words, we may lock the reference frame to a narrow section of the cylindrical resonator where the rotating-wave modulation can be approximated by a simple linear travelling-wave modulation as illustrated in Figure 2. The equivalent linear translation modulation velocity is $v_0 = R\Omega_0$ and is directed along the azimuthal direction $\hat{\boldsymbol{\varphi}}$. The equivalent linear lattice constant is $R\Delta\varphi$.

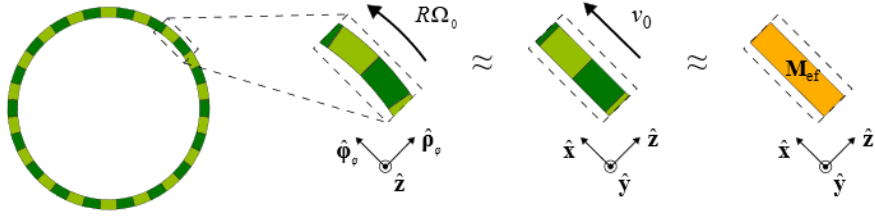

**Fig. 2** The local travelling-wave modulation approximation. For large curvature radii, each resonator section can be locally approximated by a spacetime crystal with a linear travelling-wave modulation, which in the long wavelength limit behaves as an effective medium.

According to Ref. [29], the corresponding homogenized material is characterized by uniaxial-type permittivity and permeability tensors with the optical axis along $\hat{\boldsymbol{\varphi}}$:

$$\boldsymbol{\varepsilon}_{\text{ef}} = \varepsilon_{\parallel}^{\text{ef}} \hat{\boldsymbol{\varphi}}_\varphi \otimes \hat{\boldsymbol{\varphi}}_\varphi + \varepsilon_{\perp}^{\text{ef}} \left( \hat{\boldsymbol{\rho}}_\varphi \otimes \hat{\boldsymbol{\rho}}_\varphi + \hat{\mathbf{z}} \otimes \hat{\mathbf{z}} \right). \tag{1a}$$

$$\boldsymbol{\mu}_{\text{ef}} = \mu_{\parallel}^{\text{ef}} \hat{\boldsymbol{\varphi}}_\varphi \otimes \hat{\boldsymbol{\varphi}}_\varphi + \mu_{\perp}^{\text{ef}} \left( \hat{\boldsymbol{\rho}}_\varphi \otimes \hat{\boldsymbol{\rho}}_\varphi + \hat{\mathbf{z}} \otimes \hat{\mathbf{z}} \right). \tag{1b}$$

We denote $\hat{\boldsymbol{\rho}}_\varphi = \cos(\varphi)\hat{\mathbf{x}} + \sin(\varphi)\hat{\mathbf{y}}$ and $\hat{\boldsymbol{\varphi}}_\varphi = -\sin(\varphi)\hat{\mathbf{x}} + \cos(\varphi)\hat{\mathbf{y}}$ to show explicitly the dependence of these unit vectors on the angular coordinate $\varphi$. In addition, the effective medium is characterized by bianisotropic tensors of the form $\boldsymbol{\xi}_{\text{ef}} = -\boldsymbol{\zeta}_{\text{ef}} = -\xi^{\text{ef}} \hat{\boldsymbol{\varphi}}_\varphi \times \mathbf{1}$, that is:

$$\boldsymbol{\xi}_{\text{ef}} = \xi^{\text{ef}} \left( \hat{\mathbf{z}} \otimes \hat{\boldsymbol{\rho}}_\varphi - \hat{\boldsymbol{\rho}}_\varphi \otimes \hat{\mathbf{z}} \right). \tag{1c}$$

The corresponding constitutive relations are:

$$\begin{pmatrix} \mathbf{D} \\ \mathbf{B} \end{pmatrix} = \underbrace{\begin{pmatrix} \boldsymbol{\varepsilon}_{\text{ef}}(\varphi) & \boldsymbol{\xi}_{\text{ef}}(\varphi) \\ \boldsymbol{\zeta}_{\text{ef}}(\varphi) & \boldsymbol{\mu}_{\text{ef}}(\varphi) \end{pmatrix}}_{\mathbf{M}_{\text{ef}}(\varphi)} \cdot \begin{pmatrix} \mathbf{E} \\ \mathbf{H} \end{pmatrix}. \tag{2}$$

Interestingly, due to the rotating nature of the spacetime modulation the optical axes of the effective medium change along the perimeter of the ring, and thereby the ring response is effectively inhomogeneous.

In the above, the permittivity and permeability elements in $\hat{\boldsymbol{\varphi}}_\varphi \otimes \hat{\boldsymbol{\varphi}}_\varphi$ are [29]:

$$\varepsilon_\parallel^{\text{ef}} = \left[\frac{1}{\Delta\varphi}\int_0^{\Delta\varphi}\frac{1}{\varepsilon(\varphi)}d\varphi\right]^{-1}, \qquad \mu_\parallel^{\text{ef}} = \left[\frac{1}{\Delta\varphi}\int_0^{\Delta\varphi}\frac{1}{\mu(\varphi)}d\varphi\right]^{-1}. \qquad (3)$$

The remaining permittivity, permeability and magneto-electric tensor elements are expressed in terms of the effective parameters ($\langle\varepsilon'_\perp\rangle$, $\langle\mu'_\perp\rangle$ and $\langle\xi'\rangle$) in a frame co-moving with the spacetime crystal as follows [29]:

$$\varepsilon_\perp^{\text{ef}} = \frac{\langle\varepsilon'_\perp\rangle}{\left(1+R\Omega_0\langle\xi'\rangle\right)^2 - R^2\Omega_0^{\;2}\langle\varepsilon'_\perp\rangle\langle\mu'_\perp\rangle}, \qquad (4a)$$

$$\mu_\perp^{\text{ef}} = \frac{\langle\mu'_\perp\rangle}{\left(1+R\Omega_0\langle\xi'\rangle\right)^2 - R^2\Omega_0^{\;2}\langle\varepsilon'_\perp\rangle\langle\mu'_\perp\rangle}, \qquad (4b)$$

$$\xi^{\text{ef}} = \frac{\left(1+R\Omega_0\langle\xi'\rangle\right)\langle\xi'\rangle - R\Omega_0\langle\varepsilon'_\perp\rangle\langle\mu'_\perp\rangle}{\left(1+R\Omega_0\langle\xi'\rangle\right)^2 - R^2\Omega_0^{\;2}\langle\varepsilon'_\perp\rangle\langle\mu'_\perp\rangle}. \qquad (4c)$$

The effective parameters in the co-moving frame are

$$\langle\varepsilon'_\perp\rangle = \frac{1}{\Delta\varphi}\int_0^{\Delta\varphi}\frac{\varepsilon(\varphi)}{1-\varepsilon(\varphi)\mu(\varphi)R^2\Omega_0^{\;2}}d\varphi, \qquad (5a)$$

$$\langle\mu'_\perp\rangle = \frac{1}{\Delta\varphi}\int_0^{\Delta\varphi}\frac{\mu(\varphi)}{1-\varepsilon(\varphi)\mu(\varphi)R^2\Omega_0^{\;2}}d\varphi, \qquad (5b)$$

$$\langle\xi'\rangle = \frac{1}{\Delta\varphi}\int_0^{\Delta\varphi}\frac{\varepsilon(\varphi)\mu(\varphi)R\Omega_0}{1-\varepsilon(\varphi)\mu(\varphi)R^2\Omega_0^{\;2}}d\varphi. \qquad (5c)$$

The linear travelling-wave approximation is valid provided the angular lattice period $\Delta \varphi = 2\pi / N$ satisfies $\Delta \varphi \ll 1$, or equivalently $N \gg \pi$. Since the condition depends solely on $N$ and not on $R$, this approximation can be used for sufficiently small structures, although the practical implementation for a large $N$ may be challenging.

As discussed in Ref. [29], the bianisotropic term $\xi^{\text{ef}}$ vanishes when only one constitutive parameters ($\varepsilon$ or $\mu$) is modulated. Thus, it is essential to modulate simultaneously the permittivity and permeability to break the reciprocity of the effective medium model.

Figure 1b shows the relevant effective parameters of the ring resonator for transverse electric (TE) waves ($\mathbf{E} = E_z(x, y)\hat{\mathbf{z}}$) for different values of $R\Omega_0$. The ring has a stratified geometry with two layers characterized by $(\varepsilon_{r,1} = 14, \mu_{r,1} = 1)$ and $(\varepsilon_{r,2} = 8, \mu_{r,2} = 2)$. We choose a large material contrast between layers to make more evident the impact of the spacetime modulation (the topology of the spacetime crystals considered in the following sections is independent of the strength of the material contrast). The volume fraction of the layers is identical, i.e., $f_1 = f_2 = 0.5$. The transition between the sub-luminal and super-luminal regimes $(0.25c \leq |R\Omega_0| \leq 0.267c)$ is not smooth: while in the sub-luminal regime the cylindrical shell has a regular bianisotropic response (with the material matrix positive definite), in the super-luminal regime both the effective permittivity and permeability can be negative (material matrix is indefinite). The cylindrical shell tends to an isotropic response as the rotation velocity approaches infinity.

## III. Band structure and topology of the spacetime crystal

Consider now a honeycomb lattice of spacetime modulated ring resonators. For now, each unit cell is formed by two rings embedded in a homogeneous and isotropic host medium, as shown in Figure 3. The spacetime modulated resonators are modeled using the nonuniform and bianisotropic constitutive relations (2).

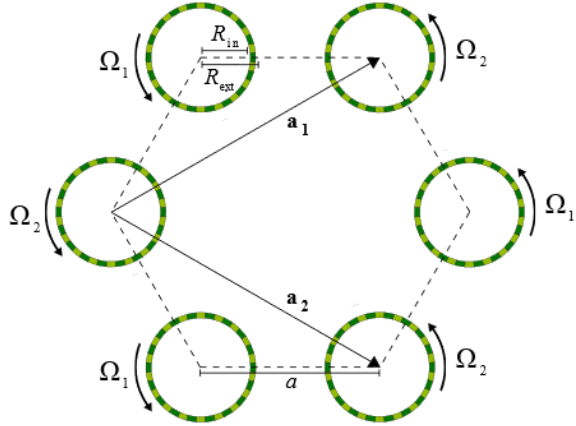

**Fig. 3** Geometry of the photonic crystal: honeycomb lattice of cylindrical resonators with a rotating spacetime modulation. Each unit cell contains two resonators subject to the angular velocities $\Omega_1$ and $\Omega_2$.

We are interested in E-polarized waves (TE, with respect to the *xoy* plane) such that $\mathbf{E} = E_z(x,y)\hat{\mathbf{z}}$ and $\mathbf{H} = H_x(x,y)\hat{\mathbf{x}} + H_y(x,y)\hat{\mathbf{y}}$. Importantly, due to the homogenization, the effective response of the rings is effectively time independent (see Eq. (2)). The effects of the time modulation are indirectly taken into account in the effective parameters. Within these approximations, for a constitutive relation of the type (2) the Maxwell's equations reduce to:

$$\hat{L} \cdot \mathbf{\Psi} = i \frac{\partial}{\partial t} \mathbf{M}(\mathbf{r}) \cdot \mathbf{\Psi}, \qquad \hat{L} = \begin{pmatrix} 0 & -i\partial_y & i\partial_x \\ -i\partial_y & 0 & 0 \\ i\partial_x & 0 & 0 \end{pmatrix}, \qquad (6)$$

where $\Psi = \begin{pmatrix} E_z & H_x & H_y \end{pmatrix}^T$ is a three-component state vector determined by the electromagnetic fields, $\hat{L}$ is a differential operator and $\mathbf{M}$ is a reduced material matrix determined by the effective parameters of the homogenized ring:

$$\mathbf{M}(\mathbf{r}) = \begin{pmatrix} \varepsilon_{zz}(\mathbf{r}) & \xi_{zx}(\mathbf{r}) & \xi_{zy}(\mathbf{r}) \\ \zeta_{xz}(\mathbf{r}) & \mu_{xx}(\mathbf{r}) & \mu_{xy}(\mathbf{r}) \\ \zeta_{yz}(\mathbf{r}) & \mu_{yx}(\mathbf{r}) & \mu_{yy}(\mathbf{r}) \end{pmatrix}. \tag{7}$$

The material matrix of the background host material is diagonal ($\xi_{zx} = \xi_{zy} = 0 = \zeta_{xz} = \zeta_{yz}$) and reduces to $\mathbf{M} = \mathbf{M}_b = \mathrm{diag}(\varepsilon_b, \mu_b, \mu_b)$.

It is convenient to express the material matrix of the spacetime crystal in terms of the indicator functions $\chi_i(\mathbf{r})$ of the $i$-th inclusion in the unit cell ($i=1,2$). Specifically, the $i$-th indicator function is defined such that $\chi_i(\mathbf{r})=1$ in the rings associated with the $i$-th sublattice of the honeycomb structure and $\chi_i(\mathbf{r})=0$ elsewhere. The material matrix can be written as:

$$\mathbf{M}(\mathbf{r}) = \mathbf{M}_b + \sum_{i=1,2} \chi_i(\mathbf{r}) \left( \mathbf{M}_{c,i}(\varphi_i) - \mathbf{M}_b \right). \tag{8}$$

The matrix $\mathbf{M}_{c,i}(\varphi_i)$ is constructed from Eq. (2) (e.g., $\xi_{zx} = \hat{\mathbf{z}} \cdot \boldsymbol{\xi}_{ef}(\varphi_i) \cdot \hat{\mathbf{x}}$) with $\varphi_i = \varphi_i(\mathbf{r}) = \arg(\mathbf{r} - \mathbf{r}_{0,i})$ and $\mathbf{r}_{0,i}$ the center of the $i$-th ring in the unit cell.

## A.  *Band structure*

As previously mentioned, in the effective medium description the material parameters are time independent. Thereby, the natural modes of the photonic crystal are regular Bloch waves with a time variation of the type $e^{-i\omega t}$. For Bloch modes, Eq. (6) reduces to $\hat{L} \cdot \Psi = \omega \mathbf{M}(\mathbf{r}) \cdot \Psi$. Using a plane wave expansion this secular equation can be reduced to a generalized eigenvalue problem of the type:

$$\underbrace{\begin{pmatrix} [0] & \left[(\mathbf{G}_{\mathbf{p}_i} \cdot \hat{\mathbf{y}})\delta_{ij}\right] & \left[-(\mathbf{G}_{\mathbf{p}_i} \cdot \hat{\mathbf{x}})\delta_{ij}\right] \\ \left[(\mathbf{G}_{\mathbf{p}_i} \cdot \hat{\mathbf{y}})\delta_{ij}\right] & [0] & [0] \\ \left[-(\mathbf{G}_{\mathbf{p}_i} \cdot \hat{\mathbf{x}})\delta_{ij}\right] & [0] & [0] \end{pmatrix}}_{\mathbf{L_k}} \cdot \begin{pmatrix} (E_{z,i}) \\ (H_{x,i}) \\ (H_{y,i}) \end{pmatrix} = \omega \underbrace{\begin{pmatrix} \left[\varepsilon_{zz,\mathbf{p}_i-\mathbf{p}_j}\right] & \left[\xi_{zx,\mathbf{p}_i-\mathbf{p}_j}\right] & \left[\xi_{zy,\mathbf{p}_i-\mathbf{p}_j}\right] \\ \left[\zeta_{xz,\mathbf{p}_i-\mathbf{p}_j}\right] & \left[\mu_{xx,\mathbf{p}_i-\mathbf{p}_j}\right] & \left[\mu_{xy,\mathbf{p}_i-\mathbf{p}_j}\right] \\ \left[\zeta_{yz,\mathbf{p}_i-\mathbf{p}_j}\right] & \left[\mu_{yx,\mathbf{p}_i-\mathbf{p}_j}\right] & \left[\mu_{yy,\mathbf{p}_i-\mathbf{p}_j}\right] \end{pmatrix}}_{\mathbf{M}} \cdot \begin{pmatrix} (E_{z,i}) \\ (H_{x,i}) \\ (H_{y,i}) \end{pmatrix}$$
(9)

Each sub-block $\left[A_{ij}\right]$ of the matrices is an infinite-dimensional matrix with $A_{ij}$ as the $(i,j)$-th element, whereas each sub-block $(F_i)$ of the column vector represents the Fourier coefficients of the generic field $F = E_z, H_x, H_y$, such that $F = \sum_i F_i e^{i\mathbf{G}_{\mathbf{p}_i} \cdot \mathbf{r}}$ with $\mathbf{G}_{\mathbf{p}_i} = \mathbf{k} + \mathbf{G}_{\mathbf{p}_i}^0$. Here, $\mathbf{G}_{\mathbf{p}_i}^0$ is a generic point of the reciprocal lattice and $\mathbf{k}$ is the wave vector of the Bloch mode. Moreover, $\varepsilon_{zz,\mathbf{p}_i}$ are the Fourier coefficients of $\varepsilon_{zz}(\mathbf{r})$ defined such that $\varepsilon_{zz,\mathbf{p}_i} = \frac{1}{A_{cell}} \int_{cell} \varepsilon_{zz}(\mathbf{r}) e^{-i\mathbf{G}_{\mathbf{p}_i}^0 \cdot \mathbf{r}} d^2\mathbf{r}$, etc and $A_{cell}$ is the unit cell area. The infinite matrices $\mathbf{L_k}$ and $\mathbf{M}$ are Hermitian. Furthermore, $\mathbf{M}$ is positive definite in the subluminal regime. Of course, in practice the infinite-dimensional vectors and matrices must be truncated to some finite dimension $N_{max}$, so that the eigenvalue problem can be solved using numerical methods.

The Fourier coefficients of the material matrix in (9) are:

$$\varepsilon_{zz,\mathbf{q}} = \varepsilon_b \delta_{\mathbf{q0}} + \chi_\mathbf{q} \sum_{i=1,2} e^{-i\mathbf{G}_\mathbf{q}^0 \cdot \mathbf{r}_{0,i}} \left(\varepsilon_{i,\perp}^{ef} - \varepsilon_b\right). \tag{10a}$$

$$\mu_{mn,\mathbf{q}} = \hat{\mathbf{u}}_m \cdot \boldsymbol{\mu}_{\mathbf{q}} \cdot \hat{\mathbf{u}}_n, \quad m,n = x, y,$$

$$\boldsymbol{\mu}_{\mathbf{q}} = \mu_b \delta_{\mathbf{q0}} \mathbf{1}_{2\times 2} + \sum_{i=1,2} e^{-i\mathbf{G}_{\mathbf{q}}^0 \cdot \mathbf{r}_{0,i}} \left[ \left( \mu_{i,\parallel}^{\text{ef}} - \mu_b \right) \boldsymbol{\chi}_{\mathbf{q}}^{\varphi\varphi} + \left( \mu_{i,\perp}^{\text{ef}} - \mu_b \right) \boldsymbol{\chi}_{\mathbf{q}}^{\rho\rho} \right].$$  (10b)

$$\zeta_{mz,\mathbf{q}} = \xi_{zm,\mathbf{q}} = \boldsymbol{\xi}_{\mathbf{q}} \cdot \hat{\mathbf{u}}_m, \quad m = x, y,$$

$$\boldsymbol{\xi}_{\mathbf{q}} = \sum_{i=1,2} \xi_i^{\text{ef}} e^{-i\mathbf{G}_{\mathbf{q}}^0 \cdot \mathbf{r}_{0,i}} \boldsymbol{\chi}_{\mathbf{q}}^\rho$$  (10c)

where we introduced

$$\chi_{\mathbf{q}} = \frac{1}{A_{\text{cell}}} \int_{\text{cell}} \chi(\mathbf{r}) e^{-i\mathbf{G}_{\mathbf{q}}^0 \cdot \mathbf{r}} d^2\mathbf{r},$$  (11a)

$$\boldsymbol{\chi}_{\mathbf{q}}^{\rho\rho} = \frac{1}{A_{\text{cell}}} \int_{\text{cell}} \chi(\mathbf{r}) \hat{\boldsymbol{\rho}}_\varphi \otimes \hat{\boldsymbol{\rho}}_\varphi e^{-i\mathbf{G}_{\mathbf{q}}^0 \cdot \mathbf{r}} d^2\mathbf{r},$$  (11b)

$$\boldsymbol{\chi}_{\mathbf{q}}^{\varphi\varphi} = \frac{1}{A_{\text{cell}}} \int_{\text{cell}} \chi(\mathbf{r}) \hat{\boldsymbol{\varphi}}_\varphi \otimes \hat{\boldsymbol{\varphi}}_\varphi e^{-i\mathbf{G}_{\mathbf{q}}^0 \cdot \mathbf{r}} d^2\mathbf{r},$$  (11c)

$$\boldsymbol{\chi}_{\mathbf{q}}^\rho = \frac{1}{A_{\text{cell}}} \int_{\text{cell}} \chi(\mathbf{r}) \hat{\boldsymbol{\rho}}_\varphi e^{-i\mathbf{G}_{\mathbf{q}}^0 \cdot \mathbf{r}} d^2\mathbf{r}.$$  (11d)

Explicit analytical formulas for the $\chi_{\mathbf{q}}, \boldsymbol{\chi}_{\mathbf{q}}^{\rho\rho}, \boldsymbol{\chi}_{\mathbf{q}}^{\varphi\varphi}, \boldsymbol{\chi}_{\mathbf{q}}^\rho$ coefficients are given in Appendix A.

## B. Gap Chern number

It is possible to assign a topological invariant for each complete band gap of a photonic crystal known as the gap Chern number [1, 22]:

$$\mathcal{C}_{\text{gap}} = \frac{1}{2\pi} \iint_{BZ} \mathcal{F}_{\mathbf{k}} d^2\mathbf{k}.$$  (12)

Here, *BZ* stands for the first Brillouin Zone and the Berry curvature $\mathcal{F}_{\mathbf{k}} = \sum_{n \in F} \mathcal{F}_{n,\mathbf{k}}$ is the sum of the Berry curvatures of all the "filled" bands below the relevant band gap. The Berry curvature can be calculated using the Green's function approach [27, 37, 38]:

$$\mathcal{F}_{\mathbf{k}} = \frac{i}{2\pi} \int_{\omega_{\text{gap}}-i\infty}^{\omega_{\text{gap}}+i\infty} \text{Tr}\left\{\partial_{k_x} L_{\mathbf{k}} \cdot G_{\mathbf{k}} \cdot \partial_{k_y} L_{\mathbf{k}} \cdot G_{\mathbf{k}} \cdot \mathbf{M} \cdot G_{\mathbf{k}}\right\} d\omega \tag{13}$$

with the Green's function defined as $G_{\mathbf{k}} = i(L_{\mathbf{k}} - \omega \mathbf{M})^{-1}$ and $\omega_{\text{gap}}$ a frequency in the band gap. The integral in frequency is over a line parallel to the imaginary frequency axis and contained in the band gap [27, 38] ($\omega_{\text{gap}} \in ]\omega_L, \omega_U[$ with $\omega_L < \text{Re}\{\omega\} < \omega_U$ a vertical strip in the complex plane that determines the gap). It is worth mentioning that the Green's function approach can also be applied to non-Hermitian systems. In the numerical calculations the Green's function $G_{\mathbf{k}}$ is replaced by a matrix that is constructed from the operators $L_{\mathbf{k}}$, $\mathbf{M}$ obtained from the plane wave expansion method [Eq. (9)]. For more details about the numerical evaluation of the gap Chern number the reader is referred to Refs. [27, 38].

## IV.   Topological Phases

In order to illustrate the ideas and the application of the theory developed thus far, we analyze two cases of interest: in the first example, we consider a spacetime crystal formed by a honeycomb lattice of ring resonators; in the second example, we propose a spacetime analogue of the Haldane model by introducing an additional triangular sublattice of ring resonators. We restrict our analysis to the subluminal regime.

### A.   *Honeycomb Lattice*

First, let us suppose that the two ring resonators are subject to the same modulation angular velocity, i.e., $\Omega_1 = \Omega_2 \equiv \Omega_0$. The ring resonators have the same material parameters as in Fig. 1b. The inner, external and average radii of the ring resonators are $R_{in} = 0.2a$, $R_{ext} = 0.3a$ and $R = 0.25a$, where $a$ is the distance between nearest neighbors.

Figure 4 depicts the photonic band structures for the static ($R\Omega_0 = 0$, dashed lines) and "rotating" ($R\Omega_0 = 0.23c$, solid lines) cases. The band structure was calculated using the plane wave method. The plane wave expansion (9) is truncated with $N_{max} = 3$. As expected, the static case presents a Dirac cone at the high-symmetry points $K$ and $K'$. As seen in Fig. 4b (dashed lines), the spacetime modulation lifts the degeneracy at the high-symmetry points and opens a complete band gap represented by the shaded blue strip.

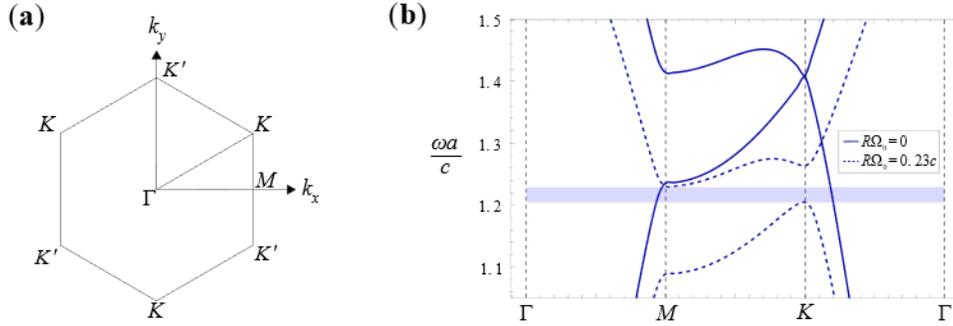

**Fig. 4 (a)** First Brillouin zone of the photonic crystal. **(b)** Band structures of the static photonic crystal ($\Omega_1 = \Omega_2 = 0$, dashed blue lines) and of the spacetime modulated crystal ($\Omega_1 = \Omega_2 \neq 0$). The angular modulation opens a band gap represented by the horizontal shaded blue strip.

To calculate the gap Chern number, we used the trapezoidal quadrature rule to numerically evaluate (12). The Brillouin zone was discretized into $N_1 \times N_2 = 25 \times 25$ points and the line integral in the complex frequency plane was truncated to an upper limit $\xi_{max} = 5.0/a^2$ and discretized into $N_\xi = 240$ points. Figure 5 shows the convergence analysis for $R\Omega_1 = R\Omega_2 = 0.23c$ and the topological phases for different pairs of modulation angular velocities in the sub-luminal regime. It is found that the gap Chern number is given by $C_{\text{gap}} = \text{sgn}(\Omega_1 + \Omega_2)$, which is identical to the sign of the average angular velocity. From the theory of Refs. [4, 30], $-C_{\text{gap}}$ may be understood as a normalized angular momentum of the edge modes at the photonic crystal boundary when it is enclosed by opaque (e.g., metallic) walls. Thus, somewhat counterintuitively, we find that the angular momentum of the edge modes has a sign *opposite* to the angular modulation speed. Furthermore, the synthetic Fresnel drag due to the spacetime modulation has a sign opposite to the sign of $\xi^{\text{ef}}$ [13, 14, 29]. From Fig. 1b, one sees that $\text{sgn}(\xi^{\text{ef}}) = -1$ in the subluminal regime, which means that the synthetic Fresnel drag direction is coincident with the angular rotation direction, but opposite to the edge mode angular momentum direction.

When $\Omega_1 = -\Omega_2 \neq 0$ there is no band gap due to the formation of two Dirac cones, despite the broken time-reversal symmetry. In this case, the topological phases are not defined.

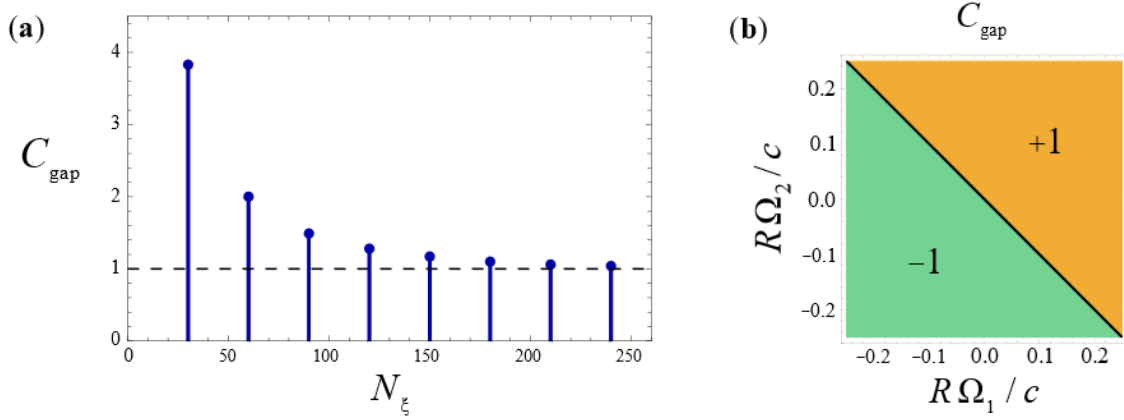

**Fig. 5 (a)** Gap Chern number convergence analysis for $R\Omega_1 = R\Omega_2 = 0.23c$ as a function of the number of points $N_\xi$ used to discretize the integral path in the complex frequency plane. **(b)** Topological phase diagram as a function of the modulation angular velocities $\Omega_1$ and $\Omega_2$.

### B. Spacetime Haldane Model

The concept of local travelling-wave modulation can be extended to a wide range of scenarios where the spacetime modulation creates a synthetic motion in arbitrary (not necessarily circular) periodic closed orbits. In that case, one can locally approximate the problem to a travelling-wave linear modulation with synthetic velocity $\mathbf{v}(\mathbf{r})$. Following the same ideas as in Sect. II, one can prove that in the long wavelength limit the resulting effective bianisotropic tensors vary in space as

$$\xi_{\text{ef}}(\mathbf{r}) = -\zeta_{\text{ef}}(\mathbf{r}) = -\xi^{\text{ef}}(\mathbf{r})\hat{\mathbf{v}}(\mathbf{r}) \times \mathbf{1}. \tag{14}$$

Here, $\hat{\mathbf{v}}(\mathbf{r})$ is the unit vector determined by the synthetic velocity. Clearly, the effective bianisotropic tensors are anti-symmetric, which corresponds to a moving-medium coupling.

Interestingly, a few recent works introduced a photonic analogue of the Haldane model [39] with the synthetic magnetic field implemented using a pseudo-Tellegen response [38, 40] (see also Ref. [41] for the corresponding electronic Haldane model in artificial graphene, and Ref. [42] for a different class of Tellegen metacrystals). The pseudo-Tellegen response is described by tensors $\xi_{ef} = \zeta_{ef}^T$ that are *symmetric* and have vanishing trace [43]. Hence, seemingly the type of coupling required to implement the synthetic magnetic field has symmetry incompatible with the symmetry provided by the rotating spacetime modulation [Eq. (14)].

The interesting observation is that TE waves cannot probe the symmetry of $\xi_{ef}(\mathbf{r})$. In fact, for TE waves with $\mathbf{E} = E_z(x, y)\hat{\mathbf{z}}$ and $\mathbf{H} = H_x(x, y)\hat{\mathbf{x}} + H_y(x, y)\hat{\mathbf{y}}$, the bianisotropic response is fully determined by the elements $\xi_{zx}, \xi_{zy}$, independent if $\xi_{ef}(\mathbf{r})$ is a symmetric or an anti-symmetric tensor. Due to this property, the pseudo-magnetic field of Refs. [38, 40] can be as well implemented using the rotating spacetime modulation [Eq. (14)].

A straightforward analysis shows that for TE-waves the pseudo-Tellegen coupling of Ref. [38] can be exactly mimicked by a rotating spacetime modulation such that (see Appendix B):

$$\xi^{ef}(\mathbf{r})\hat{\mathbf{v}}(\mathbf{r}) = \xi_0 \frac{\sqrt{3}a}{4\pi}\hat{\mathbf{z}} \times \left[\mathbf{b}_1 \sin(\mathbf{b}_1 \cdot \mathbf{R}) + \mathbf{b}_2 \sin(\mathbf{b}_2 \cdot \mathbf{R}) + (\mathbf{b}_1 + \mathbf{b}_2)\sin((\mathbf{b}_1 + \mathbf{b}_2) \cdot \mathbf{R})\right] \quad (15)$$

where $\mathbf{b}_1$ and $\mathbf{b}_2$ are the reciprocal lattice vectors of the honeycomb lattice, and $\mathbf{R} = \mathbf{r} - \mathbf{r}_c$ with $\mathbf{r}_c$ the honeycomb's lattice center [38].

The spatial distribution of the required normalized modulation "velocity" $\xi^{\text{ef}}(\mathbf{r})\hat{\mathbf{v}}(\mathbf{r})$ is shown in Fig. 6a. As seen, $\xi^{\text{ef}}(\mathbf{r})\hat{\mathbf{v}}(\mathbf{r})$ follows circular-type orbits centered at the hexagons defined by scattering centers of the honeycomb lattice. As further discussed in Appendix B, $\xi^{\text{ef}}(\mathbf{r})\hat{\mathbf{v}}(\mathbf{r})$ plays the role of a synthetic vector potential $\mathbf{A}(\mathbf{r})$ that determines the equivalent magnetic field.

An alternative and simpler synthetic vector potential with exactly the same symmetry as in Eq. (15) can be implemented by inserting a spacetime modulated ring at the center of each hexagon, as illustrated in Fig. 6b. The rings at the hexagon vertices play the role of the graphene sites of the Haldane model and thereby their material parameters are time independent ($\Omega_- = 0$). Strictly speaking, the spacetime modulated rings at the centers of the hexagons also modify the equivalent electric potential of the Haldane model, but for simplicity we ignore such a small perturbation (which may modify slightly the equivalent Haldane's tight-binding model). Note that since $\mathbf{A}(\mathbf{r}) \sim \xi^{\text{ef}}(\mathbf{r})\hat{\mathbf{v}}(\mathbf{r})$ is a periodic function the synthetic magnetic field $\nabla \times \mathbf{A}$ has zero spatial average. Furthermore, Figs. 6a and 6b show that the synthetic magnetic potential does not contribute to the nearest neighbor interactions as the contributions of adjacent hexagons (unit cells) effectively cancel out, whereas it clearly influences the next nearest neighbors coupling, similar to the Haldane model [39].

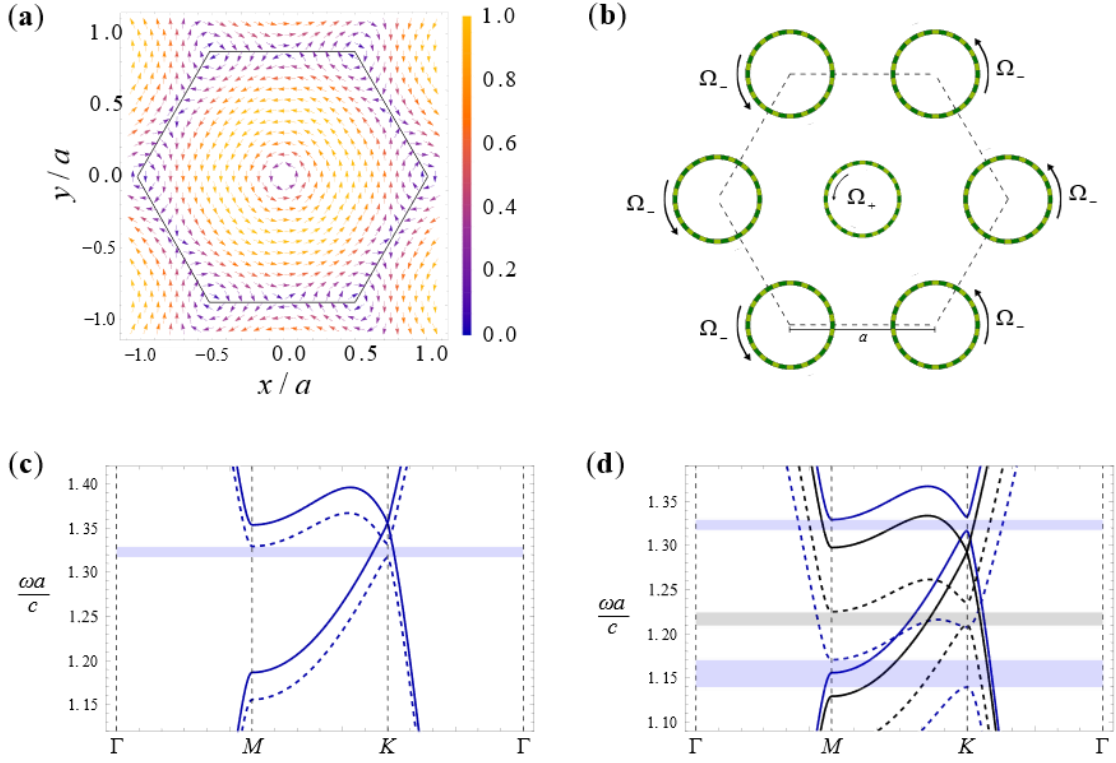

**Fig. 6 (a)** Spatial distribution of the local travelling-wave (normalized) modulation velocity $\mathbf{A}(\mathbf{r}) \sim \xi^{\text{ef}}(\mathbf{r}) \hat{\mathbf{v}}(\mathbf{r})$ required to emulate the complete spacetime analogue of the Haldane model for $\xi_0 = 1$. The unit cell region is delimited by solid black lines. **(b)** Geometry of the simplified spacetime Haldane model with the nonreciprocal coupling implemented with a spacetime modulated ring centered at the unit cell. The rings centered at the vertices of the unit cell are static ($\Omega_- = 0$). **(c)** Photonic band structure of the simplified spacetime Haldane model with the parameters $R_- = 0.25a$, $\Omega_- = 0$, $R_+ = 0.175a$ for the static ($\Omega_+ = 0$, solid blue lines) and dynamic ($\Omega_+ = 0.23c/a$, dashed blue lines) cases. The angular modulation opens a topological band gap represented by the horizontal shaded blue strip with topological charge $C_{\text{gap}} = -\operatorname{sgn}(\Omega_+)$. **(d)** Photonic band structure for different values of the modulation angular velocity of the rings centered at the vertices of the unit cell: $\Omega_- = -0.23c/a$ (dashed blue lines), $\Omega_- = 0$ (solid blue lines), $\Omega_- = 0.17c/a$ (solid blue lines), $\Omega_- = 0.22c/a$ (dashed black lines). The photonic band gaps are represented by horizontal shaded blue ($C_{\text{gap}} = -1$) and gray ($C_{\text{gap}} = +1$) strips. The structural parameters of the crystal are as in panel c) and $\Omega_+ = 0.23 c/a$.

To illustrate these ideas, next we consider a spacetime crystal formed by a static honeycomb lattice ($\Omega_- = 0$) and an additional triangular lattice of spacetime modulated ring resonators $\Omega_+ \neq 0$. The unit cell is thus formed by a resonator at the center with mean radius $R_+$ and modulation angular velocity $\Omega_+$, and by two static resonators ($\Omega_- = 0$) located at the cell's vertices (the scattering centers) with mean radius $R_-$. All the resonators have a binary structure formed by the same materials and volume fractions as in Sect. IV.A. We choose $R_- = 0.25a$, $R_+ = 0.175a$ and $\Omega_+ = 0.23 c/a$. Figure 6c presents the photonic band structures for the static ($\Omega_+ = 0$) and dynamic ($\Omega_+ \neq 0$) cases: the spacetime modulation opens a band gap with $C_{\text{gap}} = -\text{sgn}(\Omega_+)$. Curiously, different from the first example, the normalized angular momentum of the edge modes $-C_{\text{gap}}$ is now consistent with the direction of the synthetic angular Fresnel drag in the bulk region. The two examples (Sect. IV.A and Sect. IV.B) combined show that the gap Chern number sign is unrelated to the synthetic Fresnel drag direction. In particular, it follows that the topology of rotating spacetime crystals is not only a function of the average angular velocity, but it also depends on where the spacetime modulated rings are placed within the unit cell.

Any photonic crystal formed by sublattices of ring resonators with a spacetime rotating-wave modulation originates a periodic synthetic vector potential $\mathbf{A}(\mathbf{r}) \sim \xi^{\text{ef}}(\mathbf{r}) \hat{\mathbf{v}}(\mathbf{r})$. Thus, the corresponding spatially averaged synthetic magnetic field

-20-

$\langle \nabla \times \mathbf{A}(\mathbf{r}) \rangle$ is always zero. In particular, the spacetime crystals analyzed in Sect. IV.A are also associated with a pseudo-magnetic field with zero mean average.

Furthermore, the symmetry of the crystals of Sect. IV.A is also compatible with the symmetry of the pseudo-Tellegen coupling [Eq. (15)] provided $\Omega_1 = \Omega_2 = \Omega_0$. This can be understood from Fig. 6a, which shows that the pseudo-Tellegen coupling creates a synthetic magnetic potential with vortices coincident with the vertices of the hexagonal unit cell. Importantly, for $\xi_0 > 0$ the magnetic potential lines rotate in the clockwise direction near the vertices of the unit cell ($\xi^{\text{ef}} < 0$ near the vertices), whereas in the center they rotate in the opposite (anti-clockwise) direction ($\xi^{\text{ef}} > 0$ near the center). This property implies that the geometry of Sect. IV.A with $\Omega_1 = \Omega_2 = \Omega_0$ may also be regarded as an implementation of the Haldane model. Note that $\xi_0 > 0$ requires $\Omega_1 = \Omega_2 > 0$ for the geometry of Sect. IV.A and $\Omega_+ < 0$ for the geometry of this subsection, because the sign of the synthetic angular velocity is opposite to the sign of $\xi^{\text{ef}}$ for the considered ring resonator (see Fig. 1b). Due to this reason, the gap Chern numbers obtained for the two different implementations of the Haldane model are fully consistent for weak vector potentials: $C_{\text{gap}} = \text{sgn}(\xi_0)$ with $\text{sgn}(\xi_0) = \text{sgn}(\Omega_0)$ for the example of Sect. IV.A and $\text{sgn}(\xi_0) = -\text{sgn}(\Omega_+)$ for the geometry of the present section. In particular, the topological charge of the simplified spacetime implementation of the Haldane model agrees with the topological charge of the complete model [38, 40] for weak vector potentials.

It is also interesting to analyze how the topology of the spacetime crystal changes if both the rings at the vertices and the ring at the center are spacetime modulated. Figure 6d shows the photonic band structure for the same parameters as in Fig. 6c ($R_- = 0.25a$, $R_+ = 0.175a$, $\Omega_+ = 0.23c/a$) for different values of the synthetic angular velocity $\Omega_-$ of the rings centered at the unit cell's vertices. As seen, when $\Omega_- < 0$, the band gap becomes wider and $C_{gap} = -1$. This property is consistent with the fact that values of $\Omega_- < 0$ and $\Omega_+ > 0$ both contribute to realize the same topological phase (see the discussion in the previous paragraph).

In contrast, when $\Omega_- > 0$, the topological band gap ($C_{gap} = -1$) shrinks until a Dirac cone degeneracy is formed at the high-symmetry points $K$, $K'$ for a threshold value $\Omega_- = \Omega_{th} \approx 0.17c/a$. For $\Omega_- > \Omega_{th}$ the band gap reopens with $C_{gap} = +1$. In fact, when $\Omega_-$ and $\Omega_+$ have the same sign, they contribute to engineer different topological phases as the synthetic magnetic potentials associated with the resonators at the center and at the vertices whirl in opposite orientations. The topological phase transition occurs roughly for $\Omega_- = \Omega_{th} \approx 0.17c/a$, which corresponds to the angular velocity for which the spatially averaged Haldane parameter $\langle \xi_0 \rangle = 2\pi \left( R_{-,ext}^2 - R_{-,in}^2 \right) \xi_-^{ef} - \pi \left( R_{+,ext}^2 - R_{+,in}^2 \right) \xi_+^{ef} = \sum_i \alpha_i \pi \left( R_{i,ext}^2 - R_{i,in}^2 \right) \xi_i^{ef}$ vanishes. Here, $\alpha_i = 1$ if the resonators of sublattice $i$ are located at the unit cell's vertices and $\alpha_i = -1$ if they are at the unit cell's center. This approximation leads to the estimate of the gap Chern number (applicable to all the examples of Sect. IV)

$$C_{\text{gap}} = \text{sgn}\left(\sum_i \alpha_i \left(R_{i,\text{ext}}^2 - R_{i,\text{in}}^2\right) \xi_i^{\text{ef}}\right). \tag{16}$$

## V.  Conclusions

We introduced a spacetime modulated ring resonator as a basic building block for the design of topological materials. The ring resonator is subject to a rotating-wave modulation, which may be interpreted as the angular analogue of the more familiar travelling-wave modulation. We derived a simple effective medium model for the ring resonator, which describes the effect of the spacetime modulation in terms of non-homogeneous bianisotropic material parameters. Similar to Ref. [29], the bianisotropic response is nontrivial only when the two constitutive parameters $\varepsilon$ and $\mu$ are simultaneously modulated.

We applied the developed ideas to the study of the topological phases of two different spacetime crystals. The first example deals with a honeycomb array of spacetime modulated rings. Each sub-lattice of the honeycomb array is subject to a rotating-wave modulation. It was shown that the topology of the photonic crystal is controlled by the sign of the average angular modulation velocity of the ring resonators. In the second example, we introduced a simplified spacetime analogue of the photonic Haldane model [38, 40] with the synthetic magnetic potential implemented using an additional triangular sublattice of resonators. Interestingly, in this second geometry the topology induced by the spacetime modulation generally depends not only on the sign of the synthetic angular velocity of the ring resonators but also on their location within the unit cell. In both examples, a non-trivial topology arises despite the average equivalent magnetic field being zero. Our theory unveils an exciting new mechanism to engineer nontrivial material topologies without the need of an external magnetic field bias.

**Acknowledgements:** This work is supported in part by the Institution of Engineering and Technology (IET), by the Simons Foundation, and by Fundação para a Ciência e a Tecnologia and Instituto de Telecomunicações under project UIDB/50008/2020.

## Appendix A: Fourier coefficients of the material matrix

In this Appendix, we present explicit analytical formulas for the Fourier coefficients (11) of a single ring resonator centered at the origin. We start with (11a), which can be evaluated using polar coordinates:

$$\chi_{\mathbf{q}} = \frac{1}{A_{\text{cell}}} \int_{R_{\text{in}}}^{R_{\text{ext}}} r \int_{-\pi}^{\pi} e^{-i|\mathbf{G}_{\mathbf{q}}^0|r\cos(\varphi-\varphi_{\mathbf{q}})} d\varphi dr = $$
$$= \frac{2f_V}{|\mathbf{G}_{\mathbf{q}}^0|(R_{\text{ext}}^2 - R_{\text{in}}^2)} \left[ R_{\text{ext}} J_1\left(R_{\text{ext}}|\mathbf{G}_{\mathbf{q}}^0|\right) - R_{\text{in}} J_1\left(R_{\text{in}}|\mathbf{G}_{\mathbf{q}}^0|\right) \right], \quad \mathbf{q} \neq \mathbf{0} \tag{A1}$$

where $J_n(z)$ is the Bessel function of order $n$, $f_V = \dfrac{\pi(R_{\text{ext}}^2 - R_{\text{in}}^2)}{A_{\text{cell}}}$ is the volume fraction of the relevant ring and $\mathbf{G}_{\mathbf{q}}^0 = |\mathbf{G}_{\mathbf{q}}^0|(\cos\varphi_{\mathbf{q}}, \sin\varphi_{\mathbf{q}})$.

Similarly, the Fourier coefficient $\chi_{\mathbf{q}}^{\rho\rho}$ [Eq. (11b)] is given by:

$$\chi_{\mathbf{q}}^{\rho\rho} = \frac{1}{A_{\text{cell}}} \int_{R_{\text{in}}}^{R_{\text{ext}}} r \int_{-\pi}^{\pi} \hat{\boldsymbol{\rho}}_\varphi \otimes \hat{\boldsymbol{\rho}}_\varphi e^{-i|\mathbf{G}_{\mathbf{q}}^0|r\cos(\varphi-\varphi_{\mathbf{q}})} d\varphi dr$$
$$= \frac{1}{A_{\text{cell}}} \mathbf{R}_{\varphi_{\mathbf{q}}} \cdot \left[ \int_{R_{in}}^{R_{ext}} r \int_{-\pi}^{\pi} \hat{\boldsymbol{\rho}}_\nu \otimes \hat{\boldsymbol{\rho}}_\nu e^{-i|\mathbf{G}_{\mathbf{q}}^0|r\cos\nu} d\nu dr \right] \cdot \mathbf{R}_{\varphi_{\mathbf{q}}}^T \tag{A2}$$

with the rotation matrix $\mathbf{R}_{\varphi_{\mathbf{q}}}$ defined as

$$\mathbf{R}_{\varphi_{\mathbf{q}}} = \begin{pmatrix} \cos\varphi_{\mathbf{q}} & -\sin\varphi_{\mathbf{q}} \\ \sin\varphi_{\mathbf{q}} & \cos\varphi_{\mathbf{q}} \end{pmatrix}. \tag{A3}$$

After integration of (A2), one finds (for $\mathbf{q} \neq \mathbf{0}$):

$$\chi_\mathbf{q}^{\rho\rho} = -\frac{2f_V}{|\mathbf{G}_\mathbf{q}^0|^2 (R_{ext}^2 - R_{in}^2)} \left[ J_0(|\mathbf{G}_\mathbf{q}^0|R_{ext}) - J_0(|\mathbf{G}_\mathbf{q}^0|R_{in}) \right] \begin{pmatrix} -\cos 2\varphi_\mathbf{q} & -\sin 2\varphi_\mathbf{q} \\ -\sin 2\varphi_\mathbf{q} & \cos 2\varphi_\mathbf{q} \end{pmatrix}$$

$$+ \frac{2f_V}{|\mathbf{G}_\mathbf{q}^0|(R_{ext}^2 - R_{in}^2)} \left[ R_{ext} J_1(|\mathbf{G}_\mathbf{q}^0|R_{ext}) - R_{in} J_1(|\mathbf{G}_\mathbf{q}^0|R_{in}) \right] \begin{pmatrix} \cos^2 \varphi_\mathbf{q} & \cos\varphi_\mathbf{q}\sin\varphi_\mathbf{q} \\ \cos\varphi_\mathbf{q}\sin\varphi_\mathbf{q} & \sin^2 \varphi_\mathbf{q} \end{pmatrix}$$

(A4)

On the other hand, we can note that $\chi_\mathbf{q}^{\varphi\varphi}$ [Eq. (11c)] is related to $\chi_\mathbf{q}^{\rho\rho}$ in such a manner that $\chi_\mathbf{q}^{\varphi\varphi} + \chi_\mathbf{q}^{\rho\rho} = \chi_\mathbf{q} \mathbf{1}$. This implies that:

$$\chi_\mathbf{q}^{\varphi\varphi} = -\frac{2f_V}{|\mathbf{G}_\mathbf{q}^0|^2 (R_{ext}^2 - R_{in}^2)} \left[ J_0(|\mathbf{G}_\mathbf{q}^0|R_{ext}) - J_0(|\mathbf{G}_\mathbf{q}^0|R_{in}) \right] \begin{pmatrix} \cos 2\varphi_\mathbf{q} & \sin 2\varphi_\mathbf{q} \\ \sin 2\varphi_\mathbf{q} & -\cos 2\varphi_\mathbf{q} \end{pmatrix}$$

$$+ \frac{2f_V}{|\mathbf{G}_\mathbf{q}^0|(R_{ext}^2 - R_{in}^2)} \left[ R_{ext} J_1(|\mathbf{G}_\mathbf{q}^0|R_{ext}) - R_{in} J_1(|\mathbf{G}_\mathbf{q}^0|R_{in}) \right] \begin{pmatrix} \sin^2 \varphi_\mathbf{q} & -\cos\varphi_\mathbf{q}\sin\varphi_\mathbf{q} \\ -\cos\varphi_\mathbf{q}\sin\varphi_\mathbf{q} & \cos^2 \varphi_\mathbf{q} \end{pmatrix}$$

(A5)

Finally, the coefficient $\chi_\mathbf{q}^\rho$ [Eq. (11d)] may be expressed as:

$$\chi_\mathbf{q}^\rho = \frac{1}{A_{cell}} \left[ \int_{R_{in}}^{R_{ext}} r \int_{-\pi}^{\pi} \hat{\boldsymbol{\rho}}_\nu e^{-i|\mathbf{G}_\mathbf{q}^0|r\cos\nu} d\nu dr \right] \cdot \mathbf{R}_{\varphi_\mathbf{q}}^T$$

$$= \chi_\mathbf{q}^\rho (\cos\varphi_\mathbf{q}, \sin\varphi_\mathbf{q})^T, \quad \mathbf{q} \neq \mathbf{0}$$

(A6)

with

$$\chi_\mathbf{q}^\rho = \frac{-i\pi f_V}{|\mathbf{G}_\mathbf{q}^0|^2 (R_{ext}^2 - R_{in}^2)} \left[ F(R_{ext}|\mathbf{G}_\mathbf{q}^0|) - F(R_{in}|\mathbf{G}_\mathbf{q}^0|) \right],$$

(A7)

where $F(u) = uJ_1(u)H_0(u) - uJ_0(u)H_1(u)$ and $H_n(z)$ is the Struve function of order $n$.

# Appendix B: Link between the spacetime modulation and the pseudo-Tellegen material response

In this Appendix, we establish a link between the bianisotropic pseudo-Tellegen response of the photonic Haldane model [38, 40] and the anti-symmetric (moving-medium type) tensor determined by the rotating spacetime modulation [Eq. (14)].

The pseudo-Tellegen response of Refs. [38, 40] is described by

$$\boldsymbol{\xi}(\mathbf{r}) = \boldsymbol{\zeta}(\mathbf{r}) = \boldsymbol{\xi}_T(\mathbf{r}) \otimes \hat{\mathbf{z}} + \hat{\mathbf{z}} \otimes \boldsymbol{\xi}_T(\mathbf{r}) \tag{B1}$$

with $\boldsymbol{\xi}_T(\mathbf{r})$ a pseudo-Tellegen vector in the *xoy* plane whose formula is given below.

If the structure is invariant along the *z*-direction, neither the moving-medium coupling [Eq. (14)] nor the pseudo-Tellegen coupling mix the transverse electric (TE) and the transverse magnetic (TM) polarizations. Thus, the TE and TM waves are always decoupled. In particular, for TE waves, the responses obtained with the two couplings are exactly identical when the elements $\xi_{zx}, \xi_{zy}$ are the same in the two models. This can be enforced by taking,

$$\xi^{\text{ef}}(\mathbf{r})\hat{\mathbf{v}}(\mathbf{r}) = \hat{\mathbf{z}} \times \boldsymbol{\xi}_T(\mathbf{r}). \tag{B2}$$

In Refs. [38, 40], the photonic Haldane model is implemented in a honeycomb lattice of air rods embedded in a Drude metal background (photonic graphene) with a pseudo-Tellegen response characterized by the pseudo-Tellegen vector

$$\boldsymbol{\xi}_T(\mathbf{r}) = \xi_0 \frac{\sqrt{3}a}{4\pi} \Big[ \mathbf{b}_1 \sin(\mathbf{b}_1 \cdot \mathbf{R}) + \mathbf{b}_2 \sin(\mathbf{b}_2 \cdot \mathbf{R}) + (\mathbf{b}_1 + \mathbf{b}_2)\sin((\mathbf{b}_1 + \mathbf{b}_2) \cdot \mathbf{R}) \Big]. \tag{B3}$$

All the symbols are defined as in the main text. Thus, from Eq. (B2) it follows that the corresponding equivalent moving-medium coupling satisfies Eq. (15). Interestingly, $\hat{\mathbf{z}} \times \boldsymbol{\xi}_T(\mathbf{r})$ is proportional to the magnetic potential of the electronic analogue of the Haldane model introduced in Ref. [41]. Thereby, $\mathbf{A}(\mathbf{r}) \sim \xi^{\text{ef}}(\mathbf{r})\hat{\mathbf{v}}(\mathbf{r})$ may be regarded as a synthetic magnetic vector potential [40, 41].